\documentclass[traditabstract]{aa}
\pdfoutput=1
\usepackage{graphicx}
\usepackage{txfonts}
\usepackage{layouts}
\usepackage{amssymb}
\usepackage{amsmath}

\usepackage{gensymb}
\usepackage[]{natbib}
\usepackage{natbib,twoopt}
\bibpunct{(}{)}{;}{a}{}{,} %% natbib format for A&A and ApJ
\makeatletter
\nonstopmode
\newcommandtwoopt{\citeads}[3][][]{\href{http://adsabs.harvard.edu/abs/#3}%
{\def\hyper@linkstart##1##2{}%
\let\hyper@linkend\@empty\citealp[#1][#2]{#3}}}
\newcommandtwoopt{\citepads}[3][][]{\href{http://adsabs.harvard.edu/abs/#3}%
{\def\hyper@linkstart##1##2{}%
\let\hyper@linkend\@empty\citep[#1][#2]{#3}}}
\newcommandtwoopt{\citetads}[3][][]{\href{http://adsabs.harvard.edu/abs/#3}%
{\def\hyper@linkstart##1##2{}%
\let\hyper@linkend\@empty\citet[#1][#2]{#3}}}
\newcommandtwoopt{\citeyearads}[3][][]%
{\href{http://adsabs.harvard.edu/abs/#3}
{\def\hyper@linkstart##1##2{}%
\let\hyper@linkend\@empty\citeyear[#1][#2]{#3}}}
\makeatother
\usepackage[colorlinks=true, linkcolor=blue, citecolor=blue,
urlcolor=blue]{hyperref}
\usepackage{caption}
\graphicspath{{Images/}}

\begin{document}

\title{Glimpses of stellar surfaces. II. Origins of the 
photometric modulations and timing variations of KOI-1452}
\author{P. Ioannidis \and J.H.M.M. Schmitt}
\institute{Hamburger Sternwarte, Universit\"at Hamburg, Gojenbergsweg 112,
21029 Hamburg, Germany\\
\email{pioannidis@hs.uni-hamburg.de}}
\date{Received  / Accepted }
\abstract{ The deviations of the mid-transit times of an exoplanet from a 
linear ephemeris are usually the result of gravitational interactions with 
other bodies in the system. However, these types of transit timing 
variations (TTV) can also be introduced by
the influences of star spots on the shape of the transit profile. \\
Here we use the method of unsharp masking to investigate the photometric 
light curves of planets with ambiguous TTV to compare the features 
in their $O-C$ diagram with the occurrence and in-transit positions of 
spot-crossing events.  This method seems to be particularly useful for the
examination of transit light curves with only small numbers of 
in-transit data points, i.e., the long cadence light curves from {\it Kepler} 
satellite.\\
As a proof of concept we apply this method to the 
light curve and the estimated eclipse timing variations of the eclipsing binary KOI-1452, 
for which we prove their non-gravitational nature. Furthermore, we use 
the method to study the rotation properties of the primary star of the 
system KOI-1452 and show that the spots responsible for the timing variations rotate 
with different periods than the most prominent periods of the system's light curve.
We argue that the main contribution in the measured photometric variability 
of KOI-1452 originates in g-mode oscillations, which makes the primary star 
of the system a $\gamma$-Dor type variable candidate.}

\titlerunning {Photometric variations and TTV in KOI-1452 }
\maketitle

\section{Introduction}

One of the most efficient ways to study the physical parameters of an exoplanet is by studying their transits in front of their 
parent star. As shown by \citetads{2005MNRAS.359..567A}, it is 
even possible to infer the existence of additional low-mass and otherwise unobservable companions with the measurement of the mid-transit times and their deviations from linear period ephemeris.  
Depending on the actual planetary
configuration, these transit timing variations (TTV) can reach differences 
up to a few hours with respect to the linear period ephemeris of the planet, especially for systems close to low order mean motion resonances.
Although additional system components can also be inferred from precise
radial velocity (RV) measurements, the importance of the TTV method increases with the number of planets around fainter stars, where the quality
of RV measurements is decreasing. While the measurements of TTV are possible with ground-based observations, the high accuracy, continuous light curves from the {\it CoRoT} \citepads{2006ESASP1306...33B} 
and {\it Kepler} \citepads{2010ApJ...713L..79K} space missions provide
evidence for statistically significant TTV in more than 60
cases.\footnote{According to data list acquired by the web page
http://exoplanet.org } 

Despite its success, the study of exoplanets with the method of transits using high precision photometry can be somewhat problematic in the case of active host stars. 
The transit profile of planets orbiting around spotted stars become distorted when the planet is passing over dark star spots during its transit (e.g., \citetads{2009A&A...494..391R}, \citetads{2009A&A...504..561W}). 
The severity of this effect can be such that quite
large systematic errors are introduced in the calculation of the physical
parameters of the planets (see \citetads{2009A&A...505.1277C} and 
\citetads{2013A&A...556A..19O}). Furthermore,
\citetads{2016A&A...585A..72I} show that the influence of these spot-crossing 
events on the estimates of the mid-transit
times of a planet can lead to statistically significant false 
positive TTV detections under certain conditions, and
there are several cases where the detected low-amplitude TTV 
are actually believed to be caused by spot-crossing 
events (e.g., \citetads{2011ApJ...733..127S}, \citetads{2012ApJ...750..114F}, \citetads{2013A&A...553A..17S}, \citetads{2013ApJS..208...16M}). {Likewise, \citetads{2002A&A...387..969K} predict that eclipse timing variations (ETV) may be caused by spot-crossing events in the eclipses of close binaries with at least one photospherically active component (e.g., \citeads{2013ApJ...774...81T}) }

The identification of the spot-crossing events and their correlation 
with the TTV of a planet can be cumbersome, especially for planets with small periods and thus a large number of transits.  Here we show that 
the method of unsharp masking, i.e., examining the transit light curve produced by subtracting out the mean transit model (see Sect.~\ref{Sect:data_pre}) is a suitable tool to compare 
the measured $O-C$ mid-transit time variations with the transit light 
curves and to identify correlations between 
the occurrence of spot-crossing events and the measured TTV.
To demonstrate the power of this method, we present a case study of the TTV of KOI-1452 in Sect.~\ref{Sect:koi1452cs}, and finally we discuss our results in Sect.~\ref{sect:1542disc}, and 
summarize with our conclusions in sect.~\ref{Sect:Conc_trj}.

\begin{figure}[t]
\begin{center}
%\vspace{-13pt}
\includegraphics[width=\linewidth]{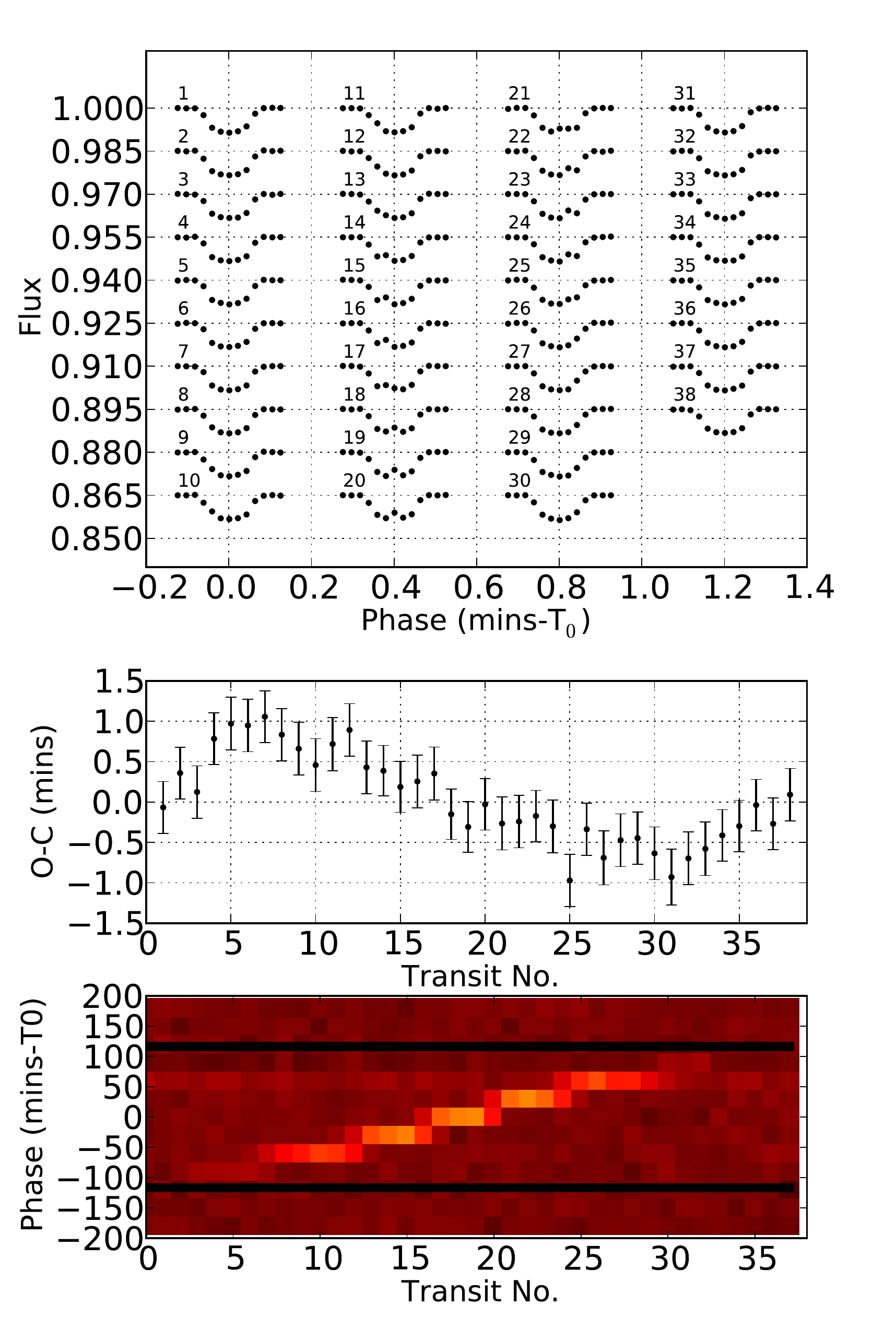}
\caption{Top: Consecutive simulated transits of a planet with an orbital period
$P_\mathrm{orb}$ = 1.0132 $P_{\star}$. Each light curve is affected by 
spot-crossing events; see text for details.\\
Middle: Estimated mid-transit times of the transits in the 
top panel. The sinusoidal shape of the $O-C$ diagram is the result of 
spot-crossing anomalies in the transits \citetads{2016A&A...585A..72I}.\\   
Bottom:  Unsharp masking transit residuals of the planet; note the correlation
with the sinusoidal O-C diagram in the middle panel.
}
\label{fig:trad_meth}
\end{center}
\end{figure}

\section{Unsharp masking of transit light curves}
\label{Sect:data_pre}

As discussed by \citetads{2016A&A...585A..72I}, 
when planets cross over cool star spots during the transit, the resulting 
transit profiles show
an increment of stellar flux compared to a pristine un-spotted transit.
Owing to the continuous evolution of starspots, these spot-induced in-transit anomalies {are usually not included} in the model fit of the transit light curve. As a result, the spot-crossing events can potentially alter the mid-transit time estimate, which is derived using a spot-free transit light curve.
The position of the spot-crossing event in the transit profile influences the amplitude and the sign of the spot-induced variation. 
The resulting maximal amplitude of this fake TTV depends on the physical 
properties of the star, the planet, and the spot. 

In consecutive transits of the same planet,
the position of the spot-crossing event can change 
from one transit to the next, depending on the period ratio
between the orbital period $P_\mathrm{orb}$ and rotational period 
$P_\mathrm{rot}$ of the star.  Fake TTV may result if $P_\mathrm{orb}$
and $P_\mathrm{rot}$ are not identical.
In particular, the 
time needed for a spot to cross the total transit profile is equal 
to half of the beating period and it is given by the expression
\begin{equation}
\label{eq:spotdur}
D_\mathrm{sp} =  {P_\mathrm{orb}\over{2}}\cdot \left({P_\mathrm{orb}\over P_\mathrm{rot}}-1\right)^{-1} ,
\end{equation}
assuming parallel orbital and rotational axes. 
Sometimes it is useful to express the value of 
D$_\mathrm{sp}$ from Eq.~\ref{eq:spotdur} in terms of the number of 
transits needed for a spot crossing, 
which is given by

\begin{equation}
D_\mathrm{sp\_tr} =  {{D_\mathrm{sp}}\over{P_\mathrm{orb}}}.
\label{eq:spotdur_tr}
\end{equation}

The sign of D$_\mathrm{sp}$
value is related the direction of the spot motion: If 
D$_\mathrm{sp}$>0, the spot crossing events occur
in the direction from transit ingress towards egress ($P_\mathrm{orb} > P_\mathrm{rot}$)
and vice versa. We note that this relation between the sign 
of D$_\mathrm{sp}$ and the spot motion in consecutive transits is 
valid for a prograde planetary orbit; in the case of a
retrograde orbit, the above correlation is reversed. 

To demonstrate the effects of spot-crossing events we carry out a simulation
of a sequence of transits with a spot-crossing event.
The simulation is geared towards
a typical transit signal-to-noise ratio (TSNR) and integration times 
of a long cadence {\it Kepler} light curve of a star with magnitude $K_p = 13$ mag \citepads{2010ApJ...713L.120J}.
We specifically assume a ratio between the orbital period of the planet $P_\mathrm{orb}$ and the rotational period of the star $P_\mathrm{rot}$ of
$P_\mathrm{orb}/P_\mathrm{rot}$ = 1.0132. 
We further assume a dark star spot ($R_\mathrm{sp} = 0.06\,R_\mathrm{\star}$) on the visible hemisphere and that is kept fixed on the stellar surface, which is occulted by the planet ($R_\mathrm{pl} = 0.08\,R_\mathrm{\star}$) during every consecutive
transit.  The resulting transit light curves are shown in the 
top panel of Figure~\ref{fig:trad_meth}.  Since orbital and rotational periods differ slightly, the position of the spot-crossing event slowly 
changes from one transit to the next, which leads to changes in the derived
mid-transit times shown in the middle panel of Fig.~\ref{fig:trad_meth}.

The identification of the spot-crossing events and the correlation of their occurrence and in-transit position to the $O-C$ diagram is one of the most efficient ways to
distinguish between real and spot-induced TTV. 
Although we know that the $O-C$ variations in Fig.~\ref{fig:trad_meth} are produced by spot crossing events, it is not trivial to locate the in-transit anomalies that are responsible for each one of the TTV. 
Given the correct global parameters for the transit model,
the light curve residuals, calculated by subtracting the model from the data, should appear featureless and just show un-correlated noise. Yet, when 
spot-crossing events occur, clearly visible features appear (see Fig.~\ref{fig:trad_meth}, bottom panel), which are straightforward to identify, and 
using the unsharp masking method provides  an easy juxtaposition
of the in-transit anomalies with the $O-C$ diagram; furthermore it can reveal important information regarding photospheric structures (e.g., cold spots) in the band of the star that is occulted by the planet.

\begin{figure*}[t]
\begin{center}
\includegraphics[width=\linewidth]{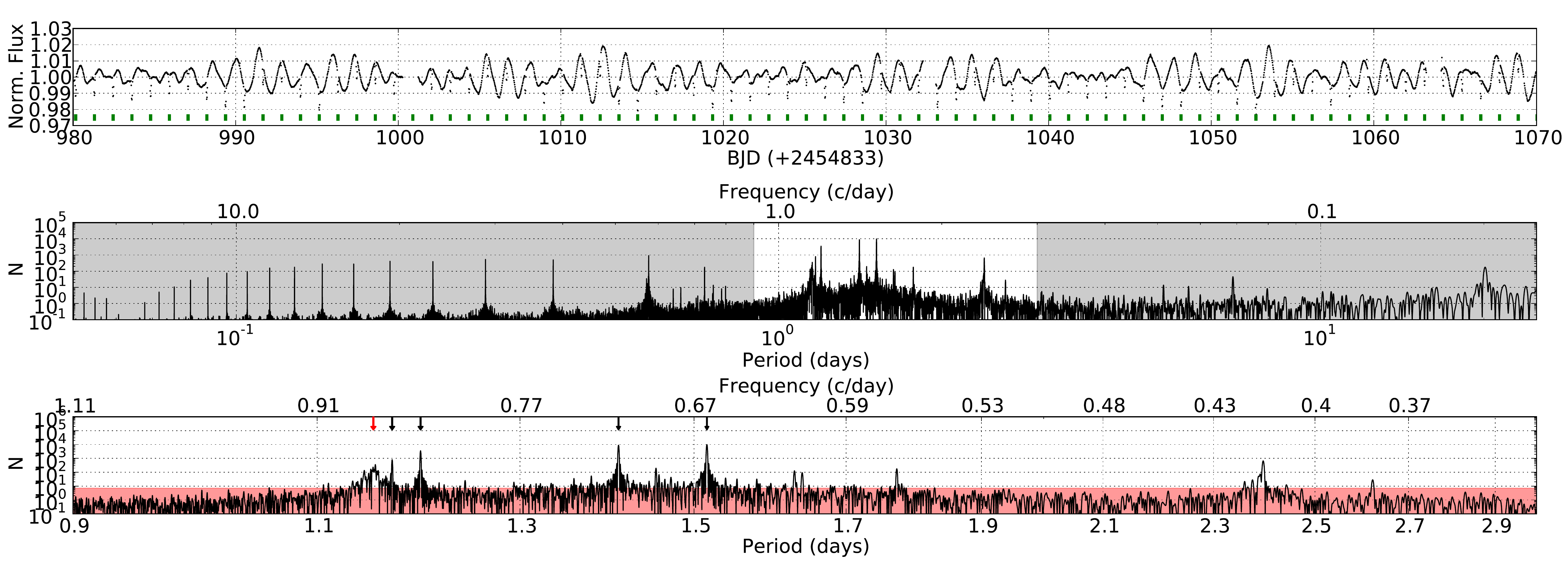}
\caption{Top: Part of KOI-1452 light curve. There are clear periodic variations,
suggesting the appearance of spots in the visible stellar hemispheres
(see text for details). The primary eclipses are indicated with green markers.\\
Middle: The L-S periodogram of the total {\it Kepler} light curve. The unshaded part around the area with the most significant periodicities  is shown enlarged in the lower panel.\\
Bottom: The red shaded area
shows the values of the periodogram with false alarm probability higher than 10$^{-3}$. The four most prominent
periods of the L-S periodogram with
periods of $P_1$~=~1.516~days, $P_2$~=~1.4097~days, $P_3$~=~1.1978~days and $P_4$~=~1.1702~days are indicated by  black arrows, while the red arrow indicates the orbital period of the system ($P_\mathrm{orb}$ = 1.1522 days).}
\label{fig:laprompeaks}
%\vspace{-10pt}
\end{center}
\end{figure*}
 
\section{Application: KOI-1452}\label{Sect:koi1452cs}
 
\begin{table}[tp]
\small
\vspace{-10pt}
\caption{System parameters for KOI-1452. }
\begin{center}
\begin{tabular}{lr}
\hline
\hline
\\[-6pt]
KIC & 7449844 \\
Kepler (mag) & 13.630\\
J (mag) & 12.727 \\
Period (days) & 1.152 $\pm$ 7$\cdot10^{-6}$$^*$  \\
R$_\mathrm{B}$/R$_\mathrm{A}$ & 0.122  $\pm$ 0.002 $^*$\\
a/R$_\mathrm{A}$ & 2.858 $\pm$ 0.012 $^*$\\
i & 73.189$\degree$ $\pm$ 0.085$\degree$$^*$\\
L$_\mathrm{B}$/L$_\mathrm{A}$ & $7.14\times10^{-4}  \pm 3.53\times10^{-4}$$^*$\\
M$_\mathrm{A}$ ($M_\mathrm{\odot}$)& 1.6 $\pm$ 0.3$^*$\\ 
M$_\mathrm{B}$ ($M_\mathrm{\odot}$)& 0.2 $\pm$ 0.1$^*$\\ 
R$_\mathrm{A}$ ($R_\mathrm{\odot}$)& 2.0 $\pm$ 0.1 $^*$\\ 
R$_\mathrm{B}$ ($R_\mathrm{\odot}$)& 0.3 $\pm$ 0.1$^*$\\ 
a ($\mathrm{au}$)& 0.0265 $\pm$ 0.0015 $^*$\\
T$_\mathrm{eff_A}$ & 7$\,$268 $\pm$ 487 $^\dagger$ \\ 
T$_\mathrm{eff_B}$ & 3$\,$582 $\pm$ 546 $^\dagger$ \\ 
Age (Myr) & 10 - 20 $^*$\\
\hline
\end{tabular}
\end{center}
\label{tab:par-1452}
$^*$ calculated in this work\\
$^\dagger$ calculated by \citetads{2014MNRAS.437.3473A}
\vspace{-10pt}
\end{table}

\subsection{Light curve analysis}
In the following, we apply the unsharp masking technique to the
{\it Kepler} data of KOI-1452.
KOI-1452 was first listed as planet host candidate by 
the {\it Kepler} satellite survey
after the identification of a transit-like signal with a period of $P_\mathrm{orb}$~=~1.1522~days. 
We show the {\it Kepler} light curve of KOI-1452 in 
 the upper panel of Fig.~\ref{fig:laprompeaks}, which demonstrates that, in addition
to the transit-like signals (shown with ticks) the light curve of KOI-1452 
shows significant modulations of the order of $\sim$2\%, suggesting an active
star, like CoRoT-2 (see \citeads{2008A&A...482L..21A} and \citeads{2009A&A...508..901H}).  
Analyzing these modulations with a
Lomb-Scargle (L-S) periodogram \citepads{2009A&A...496..577Z}, shown in
the middle panel of Fig.~\ref{fig:laprompeaks}, we find a plethora of significant periodicities with the
most prominent peak at  $P_\mathrm{rot}\simeq$~1.516 days, and
the equally separated peaks in the frequency domain with periods smaller than one day 
that can be identified as harmonics of the orbital period $P_{orb}$.  

\begin{figure}[t]
\begin{center}
%\vspace{-10pt}
\includegraphics[width=\linewidth]{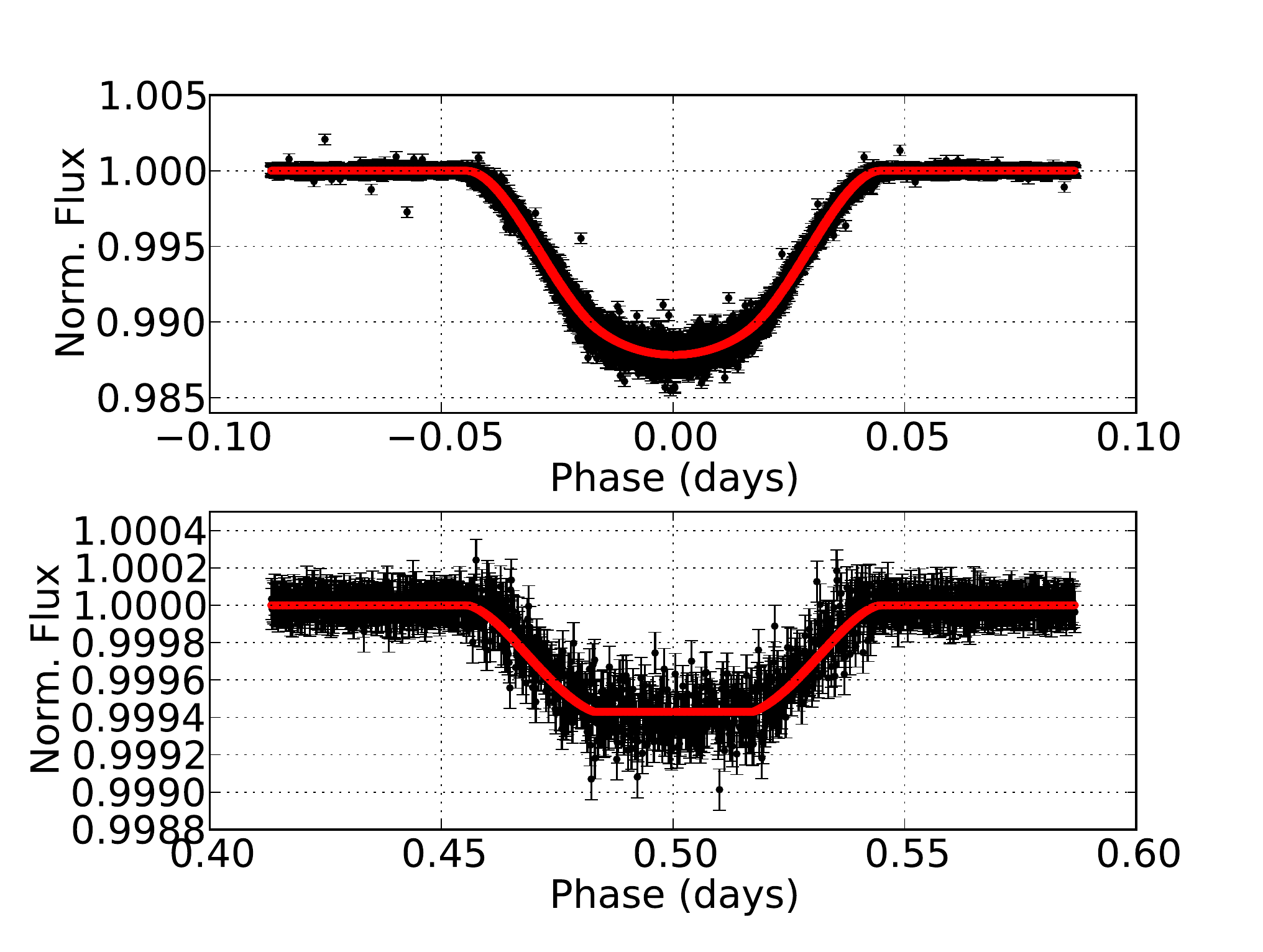}
\caption{Phase folded primary (top) and secondary (bottom) eclipses of KOI-1452 (black dots) and the 
global eclipse model (red line); see text for details.}
\label{fig:koi1452lc}
%\vspace{-10pt}
\end{center}
\end{figure}

Soon after its detection with the {\it Kepler} satellite, the KOI-1452 system 
was, however,  categorized as a stellar binary system owing to the presence of visible secondary eclipses in its light curve
\citepads{2011AJ....142..160S}. 
In Fig.~\ref{fig:koi1452lc}, we show the primary (upper panel) and secondary (lower panel)
eclipses of KOI-1452, phase folded using the calculated 
period $P_\mathrm{orb}=$1.152 days; the model shown as a red curve in Fig.~\ref{fig:koi1452lc} is constructed using the 
analytical eclipse light curve model by \citetads{2002ApJ...580L.171M} and
a summary of our model-fit results is given in Table~\ref{tab:par-1452}. 
Clearly, the shape of the primary eclipse looks very  similar to that of a planetary transit. 
The model of the secondary eclipse (red curve in the lower panel of Fig.~\ref{fig:koi1452lc}) requires identical parameters as the model calculated from the primary eclipse, assuming an inverted flux ratio and {mid-transit times}. 
The increased dispersion of the in-transit data points is an
indication of spot-crossing events in the light curves 
(see, \citetads{2009A&A...505.1277C}). {We finally note that,
according to the studies of \citetads{2014A&A...566A.103L},  \citetads{2015AJ....150..144T} and \citetads{2015AJ....149...18K}, there is no evidence of companions in angular distances smaller than 2", therefore we do not consider 
third light contamination.}

\begin{figure}[t]
\begin{center}
%\vspace{10pt}
\includegraphics[width=\linewidth]{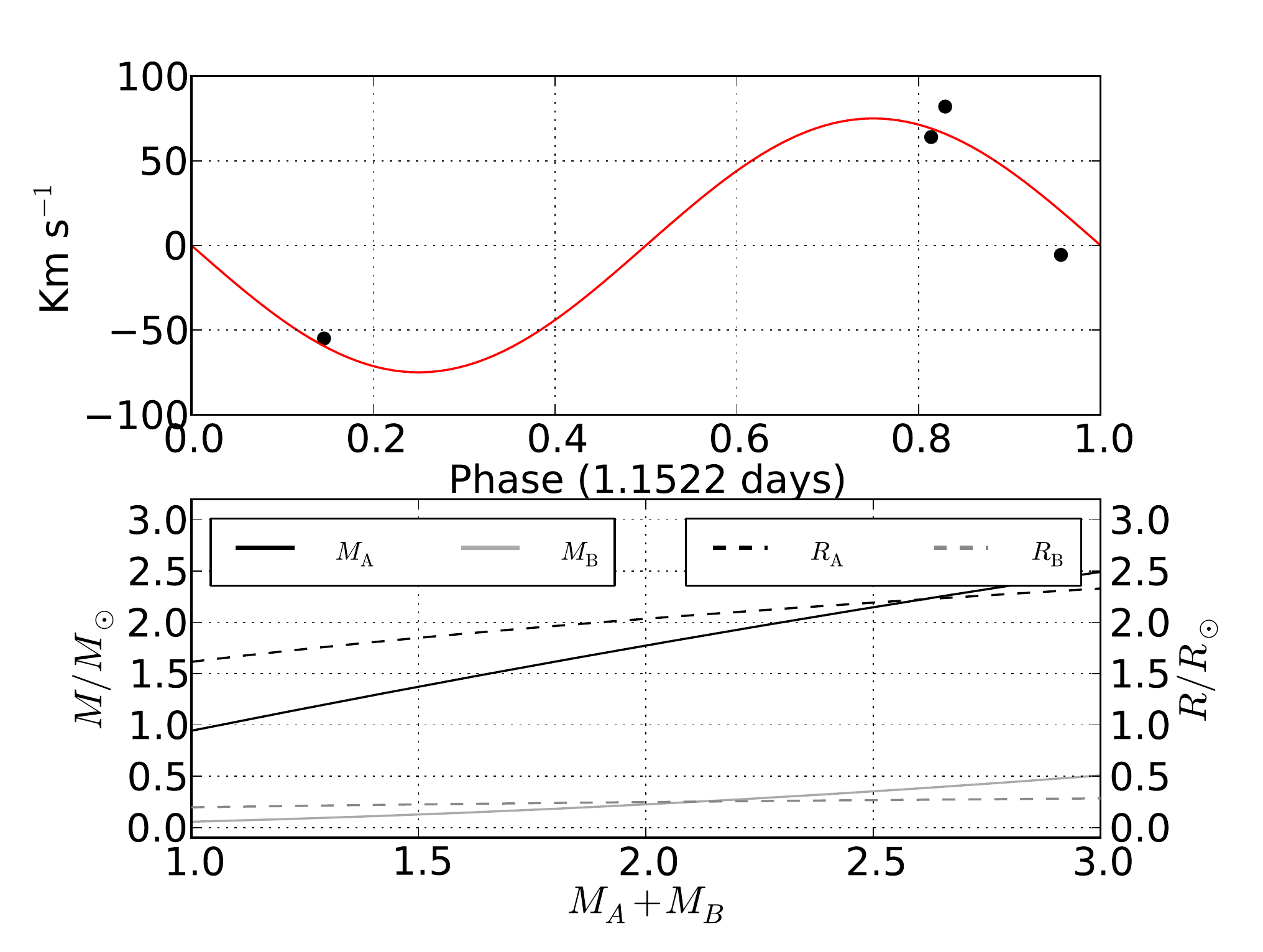}
\caption{Top: Radial velocity measurements of KOI-1452 phased with the orbital period $P_\mathrm{orb}$~=~1.1522~days. The red line shows the expected RV curve for a system of an F-type and an early M-type star,
primary eclipses occur in phase zero. The expected RV amplitude of K$\sim75\,km\,s^{-1}$
agrees well with the data, excluding the possibility for the companion to be a planet.\\
Bottom: Estimated masses (solid lines) and radii (dashed lines) for the components of the system as a function of the total mass $\mathrm{M_A+M_B}$; see text for details.}
\label{fig:rv}
%\vspace{-10pt}
\end{center}
\end{figure}

\subsection{Stellar characteristics}\label{Sect:koi1452pars}

The flat-bottom shape of the secondary eclipse (see Fig.~\ref{fig:koi1452lc})
shows that the fainter companion is totally eclipsed by the primary star.
Retrieving radial velocity (RV) data for the KOI-1452 system from the public archive CFOP 
(https://cfop.ipac.caltech.edu) and folding the data with the orbital period $P_\mathrm{orb}$~=~1.1522~days, we can construct an (admittedly sparse) RV curve for  KOI-1452,
shown in the top panel of Fig.~\ref{fig:rv}, which suggests an
RV semi-amplitude of K$\simeq75\,km\,s^{-1} $.
With this value of K we use the mass function and Kepler's 
third law in combination with the measured values for the 
orbital inclination $i$ and radius ratio $R_\mathrm{B}/R_\mathrm{A}$ (see Table~\ref{tab:par-1452})
to calculate an estimate for the components' masses and radii of the
system components as a function of the assumed total system mass. The resulting
curves are shown in the lower panel of Fig.~\ref{fig:rv}. Using the observed color index ($B-V\simeq0.4$), 
we can set a constraint for the total system
mass in the range ${1.6\,M_\odot\lesssim\,\mathrm{M}_\mathrm{A}+\mathrm{M}_\mathrm{B}\,\lesssim2.2\,M_\odot}$.
For values of total mass outside this range, the expected mass-radius combinations appear
to be unusual. {Using our estimates for the mass, the radius, and the temperature
of the system components, we are able to set 
constraints regarding the age of the system between $\sim$10~Myr and $\sim$20~Myr, using the isochrones 
derived by \citetads{2000A&A...358..593S} and \citetads{2012MNRAS.427..127B}. }

Comparing the flux reductions observed during the primary and secondary eclipse,
we calculate the ratio of the effective temperatures of the two components
as T$_\mathrm{eff_A}= 2.11$~T$_\mathrm{eff_B}$. This estimate is
in good agreement with the T$_\mathrm{eff}$ values
calculated by \citetads{2014MNRAS.437.3473A}.

Taking into account the masses and radii estimates (see, Fig.~\ref{fig:rv})
in combination with the ratio of the effective temperatures, we argue that the system 
consists of a slightly inflated F-type star and an M-dwarf star.

\subsection{False positive timing variations}\label{Sect:1452ttv}

\begin{figure}[t]
\begin{center}
\includegraphics[width=\linewidth]{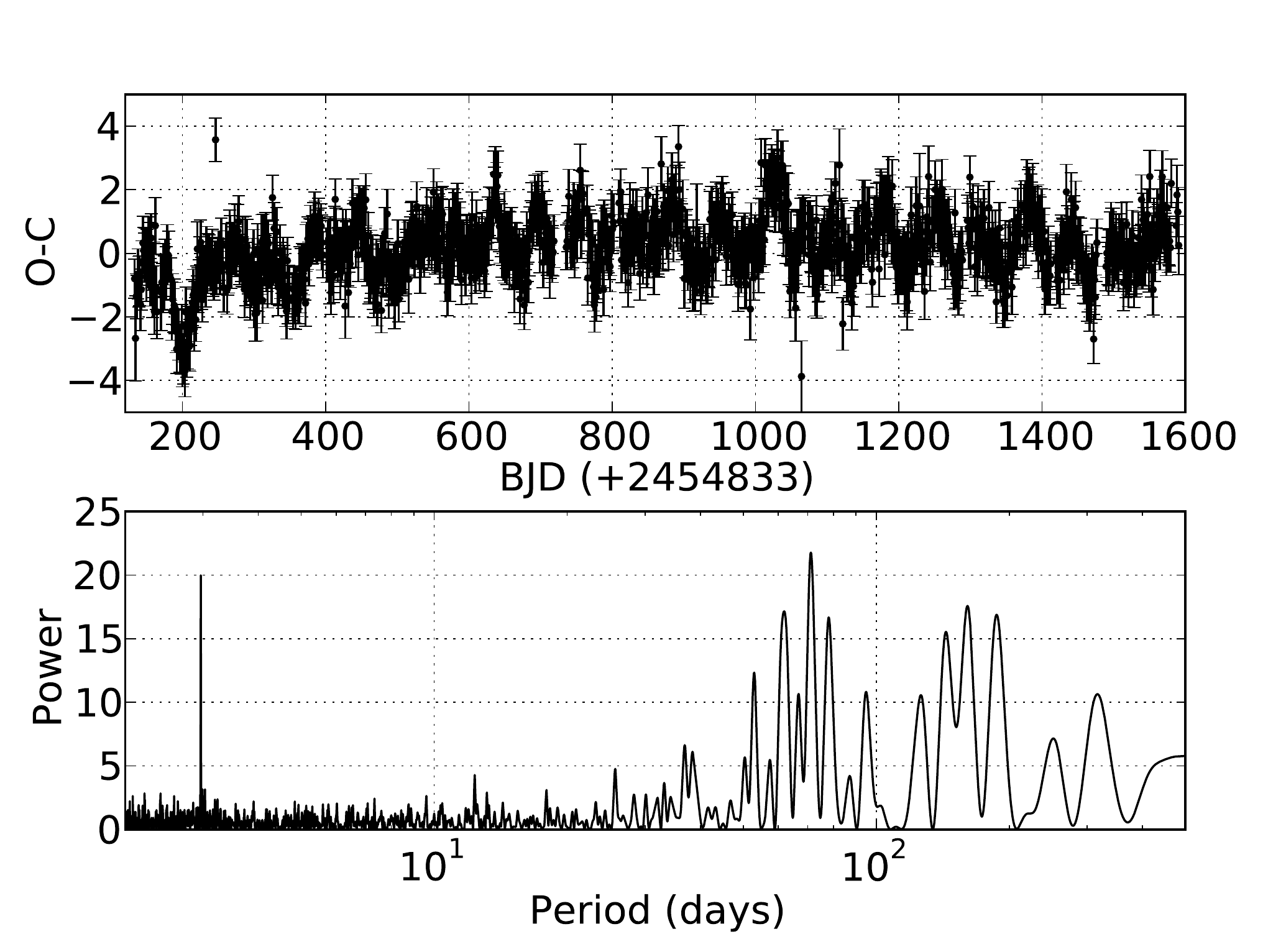}
\caption{ $O-C$ diagram of  KOI-1452 (top) and its L-S periodogram (bottom). Many significant signal periods between 60 and 100 days appear, the peak close to 3 days is caused by the data sampling.}
\label{fig:ocls1452}
%\vspace{-10pt}
\end{center}
\end{figure}

The preparation of the {\it Kepler} light curves and the ETV analysis follow the same principles described by \citetads{2016A&A...585A..72I}.  In 
the top panel of Fig.~\ref{fig:ocls1452}, we plot the $O-C$ diagram of KOI-1452\footnote{The system was introduced to us by Dr. Howard M. Relles,
http://exoplanet-science.com}, which shows clear
variations with an amplitude of $\sim$2-3 mins (\citetads{2013ApJS..208...16M} and \citetads{2013A&A...553A..17S}). 
In addition, the LS-periodogram  \citepads{2009A&A...496..577Z} 
of the derived $O-C$ values, shown in the bottom panel of Fig.~\ref{fig:ocls1452}, 
reveals a variety of high power periodicities, with the most 
prominent peak at a period $P_{O-C}\approx$71 days. The coordination of the 
rotation period of the system, in combination with the  small
number of in-transit points, periodically affect the mid-transit timing measurements; this effect 
is visible as the high peak close to $\sim$3 days in the L-S periodogram.

\begin{figure}[t]
%\vspace{10pt}
\begin{center}
\includegraphics[width=\linewidth]{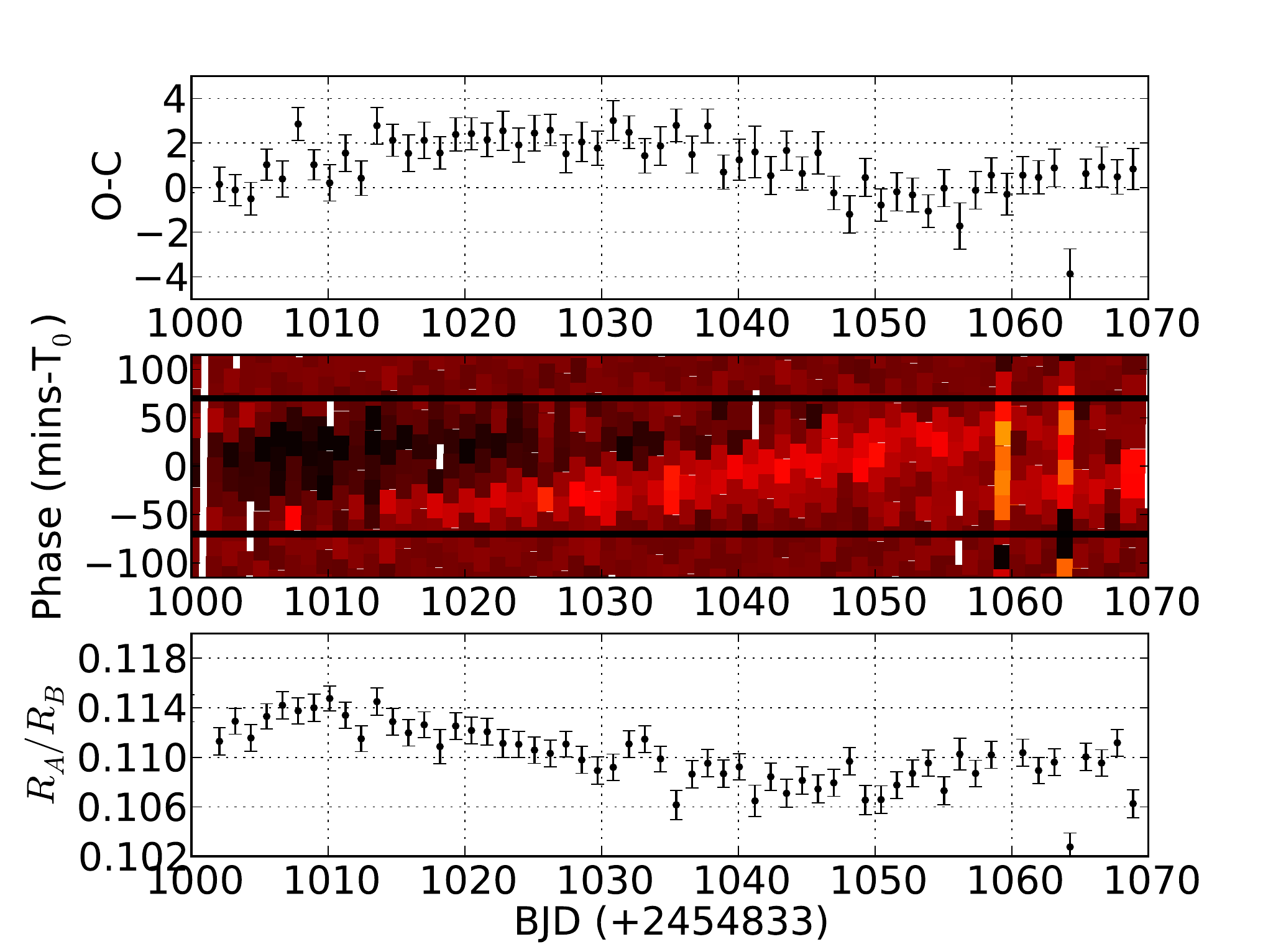}
\caption{Top: Part of $O-C$ diagram of KOI-1452 between 1000~days~$\leq$~BJD~(+2454833)~$\leq$~1070~days. \\
Middle: Concurrent unsharp masking transit light curve residuals. The bright stripe is due to spot-crossing anomalies, which change their
position from one transit to the next due to the small period difference (cf., Fig.~\ref{fig:trad_meth}). The white areas indicate no data.
The black lines indicate the first and fourth contact of the eclipse.
\\ Bottom: Variation of the 
measured radii $R_\mathrm{B}/R_\mathrm{A}$ for each eclipse. During the strongest observed TTV the star appears to be larger because of spot coverage during the transit \citep{2009A&A...505.1277C}.}
\label{fig:koi1452}
%\vspace{-10pt}
\end{center}
\end{figure}

In Fig.~\ref{fig:koi1452} we zoom in on a small portion of the $O-C$ diagram of KOI-1452 in that part of the light curve between 
1$\,$000~days~$\leq$~BJD~(+2454833)~$\leq$~1$\,$070~days 
to juxtapose the phased light curve residuals with the the $O-C$ diagram
and the derived stellar parameters.
We choose this particular part of the light curve since, in that time period, the clearest features in the $O-C$ diagram are found.  The bright, stripe-shaped feature is the result of many spot-crossing events, which 
appear in different, in-transit positions owing to the small difference between the orbital and rotational periods. A comparison 
between the $O-C$ diagram and the phased unsharp masked light curve shows that the ETV occur at the same time as those features, in good agreement with our simulations (see Fig.~\ref{fig:trad_meth}). 
The discrepancies of the $O-C$ diagram and our simulation 
are introduced by other spots, in different positions in the transit profile (e.g., a new feature 
appears in the phased color plot after BJD~(+2454833)~$\simeq$~1$\,$040~days).

As a second step, we calculate the depth for each observed 
eclipse; the relative measurements for KOI-1452 are presented in the bottom 
panel of Fig.~\ref{fig:koi1452}. According to \citetads{2009A&A...505.1277C}, 
the presence of cold spots during the transit of a planet, or in this case the eclipse of an other so-called cold star, can cause an overestimate of its radius. As a result, when the occulting body appears to be larger than average (increased transit/eclipse depth), the possibility of a 
spot-crossing event occurrence also increases. The comparison of the bottom panel of 
Fig.~\ref{fig:koi1452} with the top and middle panels shows the correlation 
of the individual occultation depths with the measured ETVs of the system and 
the existence of spot-crossing events, respectively.

Spot-crossing events are visible during the total {\it Kepler} light curve of KOI-1452, with similar characteristics and
correlation to the measured ETV and eclipse depth. Additional examples of these types of
spot-crossing events in the eclipses of KOI-1452 are shown in Fig.~\ref{fig:multi_cross}.

\section{Discussion}
\label{sect:1542disc}

\subsection{The different periods of KOI-1452}

\begin{figure}[t]
\begin{center}
\includegraphics[width=\linewidth]{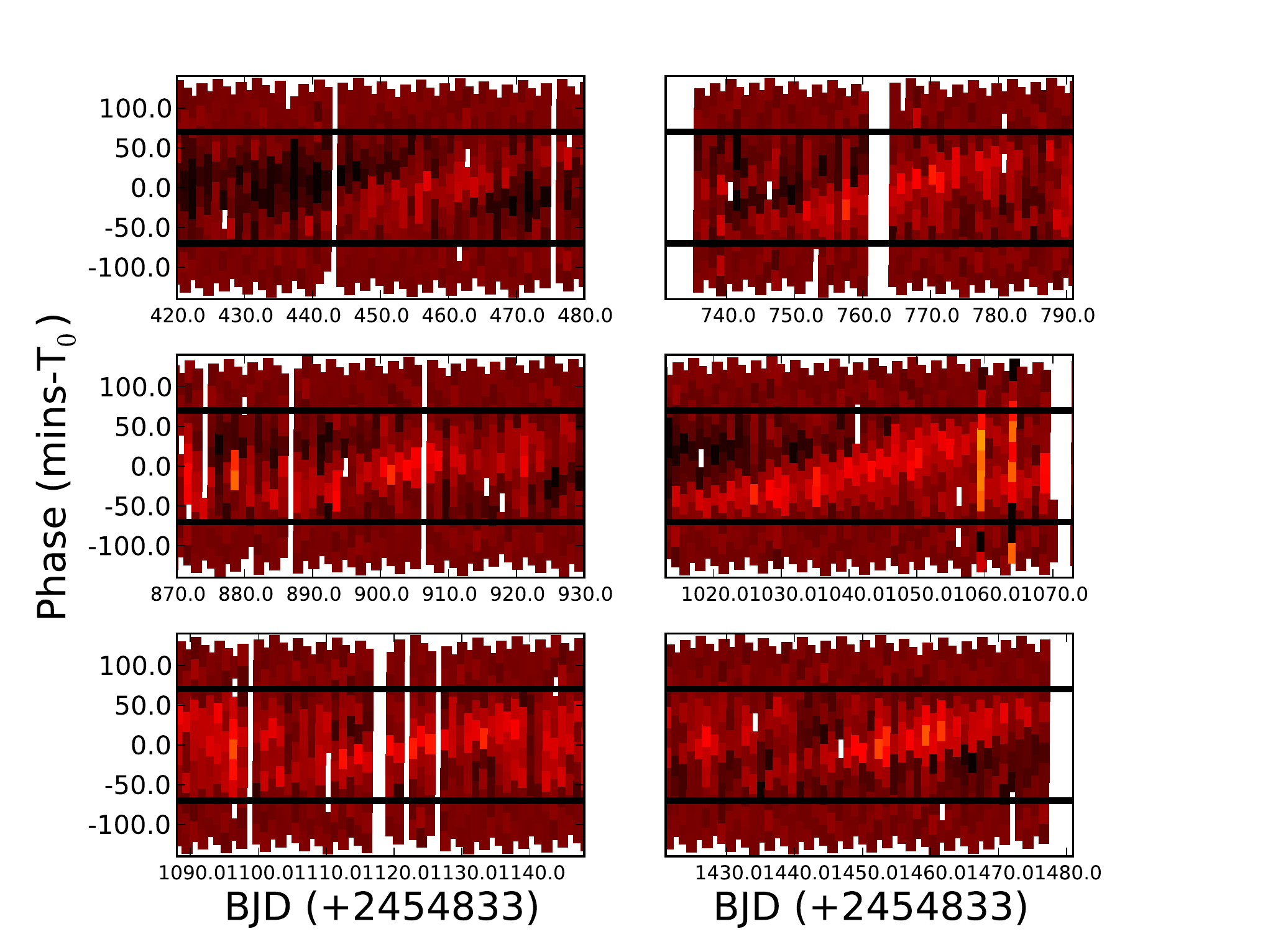}
\caption{Collection with the clearest spot-crossing features from the light curve of KOI-1452; white regions are due to
data absence. The black lines
indicate the first and fourth contact of the eclipse. }
\label{fig:multi_cross}
%\vspace{-10pt}
\end{center}
\end{figure}

As far as the observed light curve modulations of KOI-1452 are concerned, 
the most straightforward interpretation would be to attribute them to starspots and one would interpret the different peaks
in the L-S diagram as indicators of differential rotation \citepads[e.g.,][]{2013A&A...560A...4R}. {While F-type stars
 do not usually show high levels of photospheric activity, the inflated state of the KOI-1452 main component - because of its youth - allows us to consider such a possibility.} Also,
if KOI-1452 is a very active star, as suggested by its {\it Kepler} 
light curve (see Fig.~\ref{fig:koi1452lc}),
one would also expect spot-crossing events as observed, for example,
for CoRoT-2 (e.g., \citeads{2009A&A...504..561W},  \citeads{2009A&A...505.1277C} and \citeads{2010A&A...514A..39H}) and, indeed, our unsharp
masking analysis (see Figs.~\ref{fig:koi1452} and~\ref{fig:multi_cross}) does suggest spot-crossing events.
Considering the spots that are involved in spot-crossing events
and using the average of the $D_\mathrm{sp}$ 
values of the spot crossing features shown in Fig.~\ref{fig:multi_cross},
we estimate $D_\mathrm{sp}\simeq35\pm5$ days; we note that
the sign of $D_\mathrm{sp}$ is positive since we observe the spots  drifting with a direction from ingress to egress (see Sect.~\ref{Sect:data_pre}). 
Using Eq.~\ref{eq:spotdur}, we then compute the period of the spots
responsible for the spot-crossing features in the eclipses of KOI-1452 as
$P_\mathrm{rot}=\,1.1357\pm0.003$ days, assuming a prograde
orbit, or $P_\mathrm{rot}=\,1.1687\pm0.003$ days, assuming a retrograde orbit. 

{Despite the relatively young age of the system (see Table~\ref{tab:par-1452}) we estimate - using the formulation derived by \citetads{1977A&A....57..383Z} - that the synchronization and circularization processes 
should occur much earlier in the lifetime of the system (i.e., $\tau_\mathrm{circ}\lesssim$1 Myr). In combination with the light curve modeling results (i.e., the eclipses have equal durations and they are separated by half orbital period $P_\mathrm{orb}$ as shown in Fig.~\ref{fig:koi1452lc}), it is reasonable to
assume that the two stars are tidally locked}, i.e., the rotational periods
of (actually both) stars coincide with the orbital period of the system,
and thus the rotational periods ought to be close to 1.15~days.  However,
the highest peak in the LS periodogram is found at $P_\mathrm{rot}\simeq$~1.512 days (cf., Fig~\ref{fig:laprompeaks}).
If this peak was produced by star spots rotating with this period,
these spots cannot be responsible for 
the spot-crossing events observed in the transits of KOI-1452, since
for this period ratio the difference between the transit positions of the spots in consecutive transits would be much 
greater. Using Eq.~\ref{eq:spotdur_tr}, we calculate that a spot with a period of $P_\mathrm{rot}\simeq$~1.512 days requires $D_\mathrm{sp\_tr}=-1.6$ rotations to cover the total transit profile, which would lead
to two problems: first, these spots would not be 
visible continuously from one rotation to the next, resulting
in completely different ETV patterns and spot-crossing features, and second, the spots should occur with a direction from egress to ingress, opposite to the direction in which they appear to drift (see Fig.~\ref{fig:multi_cross}).

\subsection{Differential rotation~?}

\begin{figure}[t]
\includegraphics[trim = 0mm 20mm 0mm 20mm, clip,width=1\linewidth]{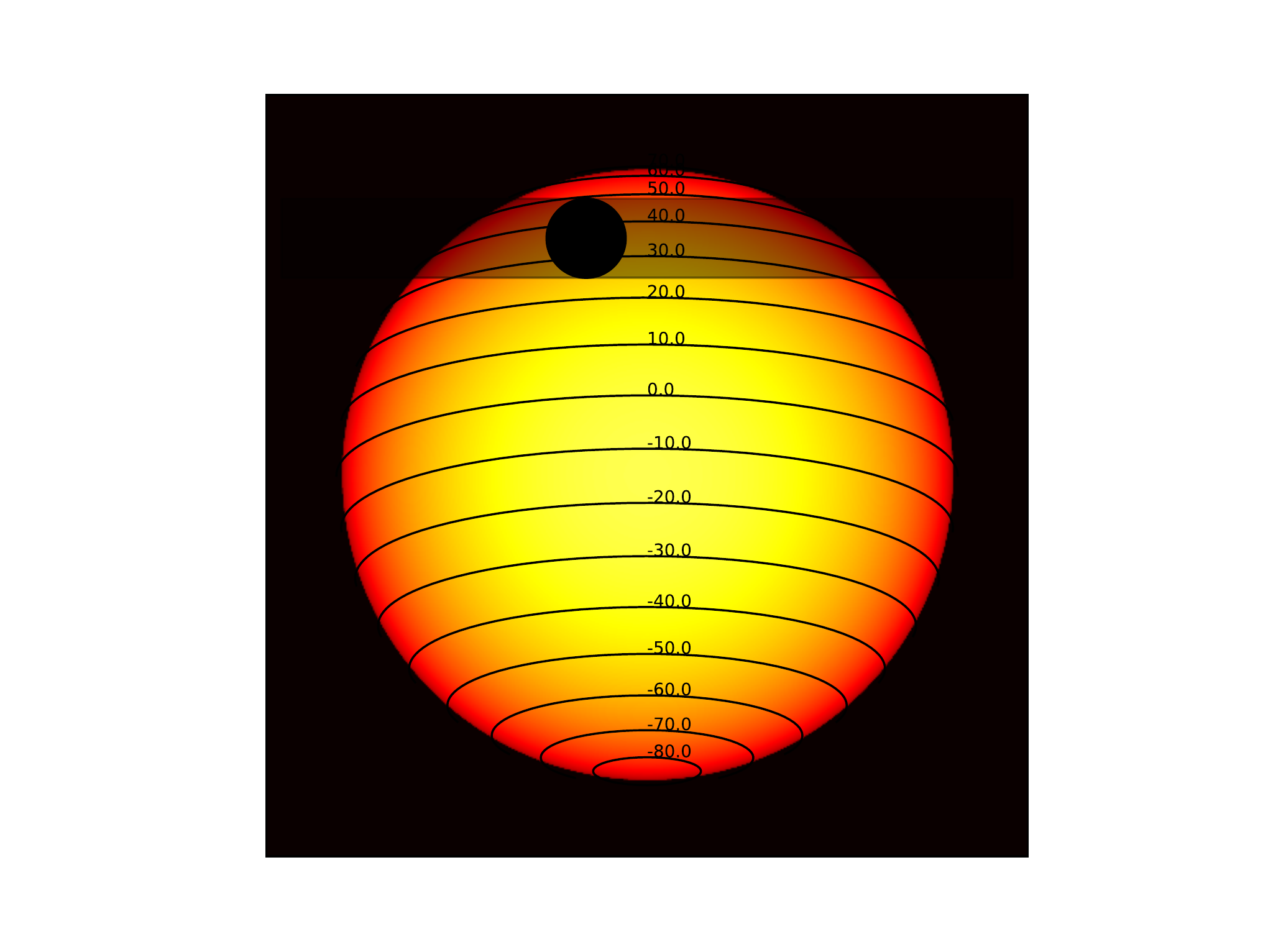}
\caption{An impression of the KOI-1452 system during a primary eclipse. The horizontal lines indicate
the latitudes of the system's primary, which appears tilted according to our hypothesis of zero system obliquity. 
The semi-transparent band represents the latitudes occulted during an eclipse. }
\label{fig:koi1452_path}
\hspace{-30pt}
\end{figure}

In Fig.~\ref{fig:koi1452_path},  we show an artist's impression of the KOI-1452 system during
a primary eclipse, using the estimated orbital characteristics shown in Table~\ref{tab:par-1452} and assuming that rotational and orbital angular momenta
are aligned. The semi-transparent black band 
represents the eclipsed part of the primary in the stellar latitude range
between $25\degree\lesssim\phi_\star\lesssim50\degree$. 
The equatorial regions at $0\degree$ are assumed to rotate with a period of
1.15~days.

It is hard to see where regions rotating with 1.51~days should be located
on the star, and also the level of differential rotation required
(period difference of $\sim$30\% between neighboring latitudes) 
is very large.
\citetads{2010MNRAS.404.1263B} discuss the possibility of zonal stellar surface flows, i.e., flows similar to those observed on the
surfaces of the giant planets in our solar system. 
Following this hypothesis would lead to accounting for the non-linear evolution of the differential rotation along the stellar latitude. As a result, 
this might explain the required rapid alternations in the rotation period.  Also, in the case of the giant planets in the solar system, the
velocity differences observed in the zonal bands are in the percent range, but not 30\%.

An even more exotic hypothesis would be that the differential rotation of the star is opposite to that of the Sun, i.e., the star rotates more slowly in the equator. Anti-solar differential rotation 
would explain the fact that the spots occulted by the companion spots, which are slower, appear 
in high latitudes. In addition, it would explain the dominance of the slow spots in the modulations
of the light curve, as they would appear closer to the equator and thus larger (as a result of the projection effect).
Although this is a plausible explanation, anti-solar differential rotation is proposed for evolved stars 
\citepads{2009ApJ...702.1078B}.

\subsection{Gravity-mode pulsations~?}

Both the area of the Hertzsprung-Russell diagram, where the main component of the KOI-1452 system is located, and
the nature of the periodicities found in the {\it Kepler} light curve, 
make it a candidate for a $\gamma$-Dor type variable star. 
These  stars are {F-type stars with effective temperatures of between $\sim$6 900 K and $\sim$7$\,$500 K, which lie on or just above the main sequence. They are} slightly more massive than the Sun (1.4-2.5$M_\odot$) and show multi-periodic oscillations with periods between 0.3 and 3 days, which are the result of gravity-mode (g-mode) pulsations \citepads[e.g.,][]{1999PASP..111..840K}. 
The mechanism responsible for the g-mode pulsations is related to the fluctuations of the
radiative flux at the base of the convective envelope of the stars \citepads{2000ApJ...542L..57G}. 
Assuming a non-rotating star with a chemically homogeneous radiative envelope, one should
expect pulsations with identical spherical degree $l$ and a variety of radial orders $n$ with 
equidistantly spaced periods \citepads{1980ApJS...43..469T}. Furthermore, stellar rotation 
introduces frequency splitting, which results in separated period spacing patterns, relative to 
the azimuthal order $m$.
\citetads{2015ApJS..218...27V} present a survey of several dozens of
$\gamma$-Dor variable stars.  While the majority of these stars
have slightly shorter pulsation periods and have period spacings
again somewhat shorter than the period spacing of P$~=~$0.1063~days
observed for KOI-1452 (see bottom panel of 
Fig.~\ref{fig:laprompeaks}), the observed parameters for  KOI-1452
do not fall completely outside the range observed for 
$\gamma$~Dor variable stars.  While an in-depth asteroseismic analysis of KOI-1452 
is beyond the scope of this paper, the hypothesis that
stellar pulsations and not stellar activity is responsible for
the observed light curve modulations of KOI-1452, appears to be
the simplest hypothesis and would not require the introduction of
somewhat exotic differential rotation scenarios.  Yet, the 
spot-crossing events visible in the primary eclipses of the
binary suggest that, in addition,  activity modulations 
must exist  which are probably described by the part of the
L-S periodogram in the period range between 1.13 days and 1.17 days.

\subsection{Retrograde orbit~?}

The assumption that the most dominant features in the L-S periodogram of 
the KOI-1452 system are the result of g-mode pulsations could explain the large discrepancy
between the measured periodicities, i.e., the difference between $P_\mathrm{orb}$ and $P_{1}$ 
in Fig.~\ref{fig:laprompeaks}. However, the fact that the spot-crossing features in the eclipses of KOI-1452
rotate with $P_\mathrm{rot}=\,1.1357\pm0.003$ days or $P_\mathrm{rot}=\,1.1687\pm0.003$ days, 
assuming a prograde or retrograde orbit respectively, requires some additional discussion.

There is no feature in the L-S periodogram close to period with value $P_\mathrm{rot}=\,1.1357\pm0.003$ days.
Yet there is a peak in the L-S periodogram at a period $P_\mathrm{rot}=\,1.1687\pm0.003$ days. Since the 
value of this peak is very close to the suggested rotational period of the occulted spots 
$P_\mathrm{rot}=\,1.1687\pm0.003$ days, it is tempting to identify the two. However, this requires us to
assume that the primary star rotates in the opposite direction of the orbital motion of its companion,
which is in contradiction to the hypothesis of tidal locking.

\section{Summary}\label{Sect:Conc_trj}

In this paper, we discuss the technique of unsharp masking as
a tool for transit light curve analysis to locate 
spot-crossing anomalies and compare their occurrence with 
the estimated $O-C$ diagram of the mid-transit times. 
We apply the method to the {\it Kepler} light curve of KOI-1452,
where we show the correlation between the timing variations and spot-crossing events in
the primary eclipses of the system.

Furthermore, we use the same method to show that it is possible to extract 
information about the properties of the spots, which are occulted during 
a transit/eclipse. As a result, we argue that the spots occulted 
by the faint companion in the KOI-1452 system are very likely not
responsible for the modulations in the {\it Kepler} light curve of the system.
We argue, furthermore, that the hypothesis that KOI-1452 is 
a $\gamma$-Dor variable can naturally explain the light curve variability
for periods of 1.19~days and longer, while star spots are responsible for
the spot-crossing events in the primary eclipses of the system and
might explain the 1.17 days feature in the periodogram.  Clearly,
ground-based observations are called for to better determine the 
physical parameters of the KOI-1452 system.  A more densely sampled RV 
curve would yield more precise estimates of the component masses, a high-resolution
spectrum would allow a better estimate of the effective temperature, and
finally, a measurement of the Rossiter-McLaughlin effect during 
primary eclipse would elucidate the rotational properties of
the primary component of  KOI-1452.

\section{Acknowledgments}

PI acknowledges funding through the DFG grant RTG
1351/2 Extrasolar planets and their host stars. In addition 
we would like to thank Sonja Schuh for her help regarding
the characterization of the primary star.

\bibliographystyle{aa}
\bibliography{references}

\newpage

\end{document}